\documentclass{PoS}

\usepackage{amssymb,amsmath,amsfonts,mathtools}
\usepackage[sort,compress]{cite}
\usepackage[bulletsep]{collref}
\usepackage{caption}
\usepackage{subcaption}
\usepackage{booktabs}

\usepackage{color}

\def \adss {$AdS_5 \times S^5$\ }
\def\la{\label}

\title{
\vspace*{-1.5cm}
{\normalsize {\rm \hfill{HU-EP-16/04} }} \\
{\normalsize {\rm \hfill{QMUL-PH-16-01} }} \\
\vspace*{1.5cm}
Lattice and string worldsheet in AdS/CFT: \\ a numerical study}
 
\ShortTitle{Lattice and string worldsheet in AdS/CFT:   a numerical study}

\author{\speaker{Valentina Forini} \\ 
        Humboldt-Universit\"at zu Berlin, Zum Gro\ss en Windkanal 6, 12489 Berlin, Germany\\
        E-mail: \email{forini@physik.hu-berlin.de}}

\author{Lorenzo Bianchi\\
        Humboldt-Universit\"at zu Berlin, Zum Gro\ss en Windkanal 6, 12489 Berlin, Germany\\
       II. Institut f\"ur Theoretische Physik, Universit\"at Hamburg, Luruper Chaussee 149, 22761 Hamburg, Germany  \\
        E-mail: \email{lorenzo.bianchi@desy.de}}

\author{Marco S.  Bianchi\\
         Center for Research in String Theory - School of Physics and Astronomy\\
        Queen Mary University of London, Mile End Road, London E1 4NS, UK\\
        E-mail: \email{m.s.bianchi@qmul.ac.uk}}
      
\author{Bjoern Leder\\
        Institutsrechenzentrum, Institut f\"ur Physik\\ 
        Humboldt-Universit\"at zu Berlin, Newtonstr. 15, 12489 Berlin, German\\
        E-mail: \email{leder@physik.hu-berlin.de}}

\author{Edoardo Vescovi\\
          Humboldt-Universit\"at zu Berlin, Zum Gro\ss en Windkanal 6, 12489 Berlin, Germany\\
        E-mail: \email{edoardo.vescovi@physik.hu-berlin.de}}                

\abstract{We consider a possible discretization for the gauge-fixed Green-Schwarz (two-dimensional) sigma-model action for  the Type IIB superstring and use it for measuring the cusp anomalous dimension of planar $\mathcal{N}=4$ SYM as derived from string theory.       
We perform lattice simulations employing  a Rational Hybrid Monte Carlo (RHMC) algorithm and a Wilson-like fermion discretization. 
 In this preliminary study, we compare our results with the expected behavior for very large values of  $g=\frac{\sqrt{\lambda}}{4\pi}$, which is the perturbative regime of the sigma-model, and find a qualitative agreement at finite lattice spacing.  
For smaller $g$ the continuum limit is obstructed by a divergence. We also detect a  phase in the fermion determinant, whose origin we explain, which 
for small $g$ leads  to a sign problem not treatable via standard reweigthing.  Results  presented here are discussed thoroughly in~\cite{toappear}.}

\FullConference{The 33rd International Symposium on Lattice Field Theory\\
		14 -18 July 2015\\
		Kobe International Conference Center, Kobe, Japan*}

\begin{document}

\section{Discussion}
\label{sec:discussion}

An impressive amount of evidence obtained in the last decade indicates the maximally supersymmetric and superconformal $\mathcal{N}=4$ planar super Yang-Mills (SYM) theory as the first example of a non-trivially interacting, four-dimensional gauge theory that might be exactly solvable. 
The ``evidence'' refers to the several results, obtained relying on the assumption of an all-order quantum integrability for this model~\cite{Beisert_review}, 
that have been confirmed by direct  perturbative  computations both in gauge theory and in what AdS/CFT assigns as string dual to $\mathcal{N}=4$ SYM - the Type IIB, Green-Schwarz string  propagating in the maximally supersymmetric background $AdS_5\times S^5$ supported by a self-dual Ramond-Ramond (RR) five-form flux. 
However, without the assumption of quantum integrability, very little is known about the theory at finite coupling. Here a lattice approach may be powerful.
A rich and interesting program of putting $\mathcal{N}=4$ SYM on the lattice is being carried out for some years by Catterall et al.~\cite{Catterall_physrept}~\footnote{We refer the reader to the POS contribution \cite{Schaich:2015ppr} for a report on interesting recent results.}. 

Following the earlier proposal of~\cite{Roiban}, we take a different lattice-based route  to investigate the finite coupling region
of the $\mathcal{N}=4$ SYM theory, discretizing the dual two-dimensional string worldsheet. We focus on a particularly important observable of the theory -- the cusp anomalous dimension -- which we describe in Section \ref{sec:observable}. 
From the dual point of view provided by the AdS/CFT correspondence~\cite{MaldaWL}, this quantity is measured by the path integral of an open string bounded by a cusped Wilson loop at the AdS boundary, which is where the four-dimensional gauge-theory lives in the holographic picture.
This string worldsheet is a highly non-trivial 2d non-linear sigma-model with rich non-perturbative dynamics.

We propose a discretization of such a model  in Section \ref{sec:discretization}. 
It is important to emphasize that, since the gauge symmetries of the model (bosonic diffeomorphisms and $\kappa$-symmetry) are all fixed, this is \emph{not} a definition  of the Green-Schwarz worldsheet string model \`a la Wilson lattice-QCD, but rather an investigation of possible routes via which lattice simulations could be an efficient tool in numerical holography.  

In  Section \ref{sec:simulation} we describe our lattice simulations, which use the Rational Hybrid Monte Carlo (RHMC) algorithm.
Our line of constant physics demands physical masses to be kept constant while approaching the continuum limit, which in the case of finite mass renormalization requires no tuning of the ``bare'' mass parameter of the theory (the light-cone momentum $P^+$).
For one of the bosonic fields entering the Lagrangian we determine the correlator and physical mass, confirming the expected finite renormalization and thus no need of tuning.  A good agreement with the perturbative sigma-model expectation  is reached  for both the correlators and the observable under investigation at very  large values of the coupling $g=\sqrt{\lambda}/(4\pi)$ ($\lambda$ is the 't Hooft coupling of the AdS/CFT dual gauge theory)~\footnote{In the AdS/CFT context, where the 't Hooft coupling $\lambda\sim g^2$ is used as relevant parameter, the large $g$ region is naturally referred to as ``strong coupling'' regime. 
The string worldsheet sigma-model of interest here, for which perturbation theory is a  $1/g$ expansion,  is however weakly-coupled at large $g$.}. For smaller $g$ we observe a divergence, possibly signaling the presence of infinite mass renormalization for the field content of the theory which so far we have not analyzed and whose study we leave for the future. 


One important result of our analysis is the detection of a phase in the fermionic determinant, resulting from integrating out the fermions. This phase  is introduced by the linearization of fermionic interactions used in \cite{Roiban}. 
For values of the coupling approaching the non-perturbative regime (corresponding to weakly-coupled $\mathcal{N}=4$ SYM) 
the phase undergoes strong fluctuations, signaling a severe sign problem. 
It would be desirable to find alternative ways to linearize quartic fermionic interactions, with resulting Yukawa terms leading to a real, positive definite fermionic determinant. Attempts in this direction are ongoing and we hope to report on them in the near future. 

\section{The observable: cusp anomaly of planar $\mathcal{N}=4$ SYM from string theory}
\label{sec:observable}

The cusp anomaly of $\mathcal{N}=4$ SYM, in this framework often simply referred to as ``scaling function''~\footnote{The  ``scaling function'' $f(g)$ is in fact the coefficient of $\log S$  in the large spin $S$  anomalous dimension $\Delta$  of leading twist operators $\Delta=f(g)\log S+{\cal O}(\log S/S)$. It  equals twice 
the cusp anomalous dimension $\Gamma_{\rm cusp}$ of light-like 
Wilson loops~\cite{Korchemsky} 
\begin{equation}\label{gammacusp}
\langle W[C_{\rm cusp}]\rangle \sim e^{-\Gamma_{\rm cusp}\,\gamma\, \ln \frac{\Lambda_{UV}}{\Lambda_{IR}} }\,,
\end{equation}
where $\gamma$ is the large, real parameter related to the geometric angle $\phi$ of the cusped Wilson loop
by $i\gamma=\phi$. The expectation value above is in fact extracted in the large imaginary $\phi$ limit.
},  is a function of the coupling and it governs the renormalization of a Wilson loop with a light-like cusp in $\mathcal{N}=4$ super Yang-Mills,
as well as the leading behavior, in the large spin $S$ regime, of the anomalous dimension of twist-two operators.
According to  AdS/CFT, any Wilson loop expectation  value 
should be represented by the  path integral of an open string ending at the AdS boundary 
\begin{equation}\label{Z_cusp}
 \langle W[C_{\rm cusp}]\rangle\equiv  Z_{\rm cusp}= \int [D\delta X] [D\delta\Psi]\, e^{- S_{\rm cusp}[X_{\rm cl}+\delta X,\delta\Psi]} = e^{-\Gamma_{\rm eff}}\equiv e^{-\frac{1}{8} f(g)\,V_2 }~.                         
\end{equation}
Above, $X_{\rm cl}=X_{\rm cl}(t,s)$ - with $t,s$ the temporal and spatial coordinate spanning the string worldsheet -  is the classical solution of the string equations of motion describing the world surface of an open string ending on a null cusp\cite{Giombi}. $S_{\rm cusp}[X+\delta X,\delta\Psi]$ is the action for field fluctuations over it -- the fields being both bosonic and fermionic string coordinates $X(t,s),~\Psi(t,s)$ -- and is reported below in equation \eqref{S_cusp} in terms of the effective bosonic and fermionic degrees of freedom remaining after gauge-fixing. Since the fluctuation Lagrangian has constant coefficients, the worldsheet volume $V_2=\int dt ds$  simply factorizes out~\footnote{As mentioned above, $f(g)$ equals twice the coefficient of the logarithmic divergence in \eqref{gammacusp}, for which the stringy counterpart should be the infinite two-dimensional worldsheet volume. The further normalization of $V_2$ with a $1/4$ factor follows the convention of~\cite{Giombi}.} in front of the function of the coupling  $f(g)$, as in the last equivalence in \eqref{Z_cusp}.  
The scaling function can be evaluated perturbatively in gauge theory ($g\ll1$), and  in  sigma-model loop expansion ($g\gg 1$) as in \eqref{cuspperturbative} below. 
Assuming all-order integrability of the spectral problem for the relevant operators and taking a thermodynamical limit of the corresponding asymptotic Bethe Ansatz, an integral equation~\cite{BES} can be derived which gives $f(g)$ exactly at each value of the coupling.

Rather than partition functions, in a lattice approach it is natural to study  vacuum expectation values. In simulating  the vacuum expectation value of the ``cusp'' action  
\begin{eqnarray}\label{vevaction}
\langle S_{\rm cusp}\rangle&=& \frac{ \int [D\delta X] [D\delta\Psi]\, S_{\rm cusp}\,e^{- S_{\rm cusp}}}{ \int [D\delta X] [D\delta\Psi]\, e^{- S_{\rm cusp}}} = -g\,\frac{d\ln Z_{\rm cusp}}{dg}\equiv g\,\frac{V_2}{8}\,f'(g)   ~,
  \end{eqnarray}
  we are therefore supposed to obtain information on the \emph{derivative} of the scaling function~\footnote{Here our analysis is different from the one in~\cite{Roiban}.  In particular,  $\langle S\rangle\sim\frac{f(g)}{V_2/2}$ only when $f(g)$ is  linear in $g$, which happens as from \eqref{cuspperturbative}   for \emph{large} $g$. }.  

In the continuum, the   \adss   superstring   action $S_{\rm cusp}$ describing quantum  fluctuations around  the null-cusp background  can be written after Wick-rotation as~\cite{Giombi}
\begin{eqnarray}\nonumber
&& \!\!\!\!\!\!\!\!\!\!\!\!
S_{\rm cusp}=g \int dt ds~ \Big\{ 
|\partial_{t}x+\textstyle{\frac{1}{2}}x|^{2}+\frac{1}{ {z}^{4}} |\partial_{s} {x}-\textstyle{\frac{1}{2}} {x}|^{2}+\left(\partial_{t}z^{M}+\frac{1}{2} {z}^{M}+\frac{i}{ {z}^{2}} {z}_{N} {\eta}_{i}\left(\rho^{MN}\right)_{\phantom{i}j}^{i} {\eta}^{j}\right)^{2}\\\label{S_cusp}
 && \!\!\!\!\!\!\!\!\!\!\!\!
+\frac{1}{ {z}^{4}}\left(\partial_{s} {z}^{M}-\textstyle{\frac{1}{2}} {z}^{M}\right)^{2}  
  +i\left( {\theta}^{i}\partial_{t}{\theta}_{i}+ {\eta}^{i}\partial_{t}{\eta}_{i}+ {\theta}_{i}\partial_{t}{\theta}^{i}+ {\eta}_{i}\partial_{t} {\eta}^{i}\right)-\textstyle{\frac{1}{{z}^{2}}}\left( {\eta}^{i}{\eta}_{i}\right)^{2}  \\\nonumber
 &&  \!\!\!\!\!\!\!\!\!\!\!\!
 +2i\Big[\textstyle{\frac{1}{z^{3}}}z^{M} {\eta}^{i}\left(\rho^{M}\right)_{ij}
 \left(\partial_{s} \theta^j-\textstyle{\frac{1}{2}} \theta^j-\frac{i}{{z}} {\eta}^{j}\left(\partial_{s} {x}-\frac{1}{2} {x}\right)\right)
\textstyle{+\frac{1}{{z}^{3}}{z}^{M}{\eta}_{i} (\rho_{M}^{\dagger} )^{ij}\left(\partial_{s}{\theta}_{j}-\frac{1}{2}{\theta}_{j}+\frac{i}{{z}}{\eta}_{j}\left(\partial_{s}{x}-\frac{1}{2}{x}\right)^{*}\right)\Big]\,\Big\}}
\end{eqnarray}
Above, $x,x^*$ are the two bosonic $AdS_5$ (coordinate) fields transverse  to the $AdS_3$ subspace of the classical solution, and $z^M\, (M=1,\cdots, 6)$ are the bosonic coordinates of the $AdS_5\times S^5$ background in Poincar\'e parametrization, with  $z=\sqrt{z_M z^M}$, remaining after fixing a ``AdS light-cone gauge''~\cite{MTMTT}. 
 The fields $\theta_i,\eta_i,\, i=1,2,3,4$ are  4+4 complex anticommuting variables for which  $\theta^i = (\theta_i)^\dagger,$ $\eta^i = (\eta_i)^\dagger$. They transform in the fundamental representation of the $SU(4)$ R-symmetry and do not carry (Lorentz) spinor indices.  The matrices $\rho^{M}_{ij} $ are the off-diagonal
blocks of $SO(6)$ Dirac matrices $\gamma^M$ in the chiral representation
and
$(\rho^{MN})_i^{\hphantom{i} j} = (\rho^{[M} \rho^{\dagger N]})_i^{\hphantom{i} j}$ are  the
$SO(6)$ generators. 
The action \eqref{S_cusp} is manifestly missing a massive parameter~\footnote{As standard in the literature - the light-cone momentum can be consistently set to the unitary value, $p^+=1$. In the perspective adopted here, however, it is crucial to keep track of dimensionful parameters as they  are in principle subject to renormalization.}, which we restore in the following defining it as $m$. 
We emphasize that, in \eqref{S_cusp},  local bosonic (diffeomorphism) and fermionic ($\kappa$-) symmetries originally present have been fixed. With this action one can directly proceed to the perturbative evaluation of the effective action in~\eqref{Z_cusp}, up to two loops in sigma-model perturbation theory~\cite{Giombi}, obtaining for the cusp anomaly  ($K$ is the Catalan constant)
\begin{flalign}\la{cuspperturbative}
f(g)=4\,g\,\Big(1-\frac{3\log2}{4\pi\,g}-\frac{K}{16\,\pi^2\,g^2}+\mathcal{O}(g^{-3})\Big)& 
\,.
\end{flalign}
Furthermore, with the same action it is possible to study perturbatively the (non-relativistic) dispersion relation for the field excitations over the classical string surface.  
For example, the corrections to the masses of the bosonic fields $x,x^*$ in \eqref{S_cusp}  (defined as the values of energy at vanishing momentum)  read~\cite{Giombi:2010bj}
\begin{equation}\label{mx}
m^2_{ x}(g)=\frac{m^2}{2}\,\Big(1-\frac{1}{8 \,g}+\mathcal{O}(g^{-2})\Big)~,
\end{equation}
where, as mentioned above, we restored the dimensionful parameter $m$.  
 In what follows, we will compute the lattice correlators of the fields $x,x^*$ so to study whether  our discretization changes the renormalization pattern above. 
%
%
\section{Linearization and discretization}
\label{sec:discretization}
While the bosonic part of~\eqref{S_cusp} can be easily discretized and simulated, Gra\ss mann-odd fields are 
formally integrated out, letting their determinant to become part -- via exponentiation in terms of pseudo-fermions, see \eqref{fermionsintegration} below -- of the Boltzmann weight of each configuration in the statistical ensemble. In the case of higher-order fermionic interactions -- as in \eqref{S_cusp}, where they are at most quartic -- this is possible via the introduction of auxiliary fields realizing a linearization. Following~\cite{Roiban}, one introduces $7$ real auxiliary  fields, one scalar $\phi$ and a  $SO(6)$ vector field $\phi_M$, with a  Hubbard-Stratonovich transformation~
\begin{eqnarray}\label{HubbardStratonovich}
&& \!\!\!\!\!\!\!
\exp \Big\{-g\int dt ds  \Big[-\textstyle{\frac{1}{{z}^{2}}}\left( {\eta}^{i}{\eta}_{i}\right)^{2}  +\Big(\textstyle{\frac{i}{ {z}^{2}}} {z}_{N} {\eta}_{i}{\rho^{MN}}_{\phantom{i}j}^{i} {\eta}^{j}\Big)^{2}\Big]\}\\\nonumber
&& 
\sim\,\int D\phi D\phi^M\,\exp\Big\{-  g\int dt ds\,[\textstyle\frac{1}{2}{\phi}^2+\frac{\sqrt{2}}{z}\phi\,\eta^2 +\frac{1}{2}({\phi}_M)^2-i\,\frac{\sqrt{2}}{z^2}\phi^M {z}_{N} \,\big(i \,{\eta}_{i}{\rho^{MN}}_{\phantom{i}j}^{i} {\eta}^{j}\big)]\Big\}~.
\end{eqnarray}
Above, in the second line we have  written  the Lagrangian for $\phi^M$ so to emphasize that it has an imaginary part,  due to the fact that the bilinear form in round brackets   is hermitian
\begin{equation}
\!
\Big(i\,\eta_i {\rho^{MN}}^i{}_j \eta^j\Big)^\dagger=-i(\eta^j)^\dagger({\rho^{MN}}^i{}_j)^*(\eta_i)^\dagger
=-i \eta_j\,{\rho^{MN}}_i{}^j\,\eta^i=i\eta_j\,{\rho^{MN}}^j{}_i\,\eta^i
\,,
\end{equation}
as follows from the properties of the $SO(6)$ generators (see for example Appendix A of \cite{Giombi}). 
Since the auxiliary vector field $\phi^M$ has real support, the  Yukawa-term for it sets \emph{a priori} a phase problem, 
the  only  question being whether the latter is treatable via standard reweighing. Below we find evidence that 
this is not the case.

After the transformation \eqref{HubbardStratonovich}, the corresponding Lagrangian reads  
\begin{eqnarray}\label{Scuspquadratic}
\!\!\!\!\!\!\!\!\!\!\!\!\!\!\!
{\cal L} = \!\textstyle {| \partial_t {x} +\!\frac{m}{2}{x} |}^2 + \!\frac{1}{{ z}^4}{| \partial_s {x} -\!\frac{m}{2}{x} |}^2
+\! (\partial_t {z}^M + \!\frac{1}{2}{z}^M )^2 +\! \frac{1}{{ z}^4} (\partial_s {z}^M -\!\frac{m}{2}{z}^M)^2+\!\frac{1}{2}{\phi}^2 +\!\frac{1}{2}({\phi}_M)^2+\!\psi^T O_F \psi\,
\label{final_continuum_L}
\end{eqnarray}
  with  $\psi\equiv({\theta}^i, { \theta}_i, {\eta}^i, {\eta}_i)$ and  
\begin{eqnarray}
\label{OF}
\!\!\!
O_F & =&\left(\begin{array}{cccc}
0 & i\partial_{t} & -\mathrm{i}\rho^{M}\left(\partial_{s}+\frac{m}{2}\right)\frac{{z}^{M}}{{z}^{3}} & 0\\
\mathrm{i}\partial_{t} & 0 & 0 & -\mathrm{i}\rho_{M}^{\dagger}\left(\partial_{s}+\frac{m}{2}\right)\frac{{z}^{M}}{{z}^{3}}\\
\mathrm{i}\frac{{z}^{M}}{{z}^{3}}\rho^{M}\left(\partial_{s}-\frac{m}{2}\right) & 0 & 2\frac{{z}^{M}}{{z}^{4}}\rho^{M}\left(\partial_{s}{x}-m\frac{{x}}{2}\right) & i\partial_{t}-A^T\\
0 & \mathrm{i}\frac{{z}^{M}}{{z}^{3}}\rho_{M}^{\dagger}\left(\partial_{s}-\frac{m}{2}\right) &\mathrm{i}\partial_{t}+A & -2\frac{{z}^{M}}{{z}^{4}}\rho_{M}^{\dagger}\left(\partial_{s}{x}^*-m\frac{{x}}{2}^*\right)
\end{array}\right)~,\\
\label{Aoperator}
A  &=&\frac{1}{\sqrt{2}{z}^{2}}{\phi}_{M}\rho^{MN} {z}_{N}-\frac{1}{\sqrt{2}{z}}{\phi}\, +\mathrm{i}\,\frac{{z}_{N}}{{z}^{2}}\rho^{MN} \,\partial_{t}{z}^{M}~.
\end{eqnarray}

The  quadratic fermionic contribution  resulting from linearization gives then formally a Pfaffian ${\rm Pf}\,O_F$, which - in order to enter the Boltzmann weight and thus  be interpreted as a probability  - should be  positive definite.
For this reason, we proceed as in~\cite{Roiban} 
\begin{equation}\label{fermionsintegration}
 \int \!\! D\Psi~ e^{-\textstyle\int dt ds \,\Psi^T O_F \Psi}={\rm Pf}\,O_F\equiv(\det O_F\,O^\dagger_F)^{\frac{1}{4}}= \int \!\!D\xi D\bar\xi\,e^{-\int dt ds\, \bar\xi(O_FO^\dagger_F)^{-\frac{1}{4}}\,\xi}~,
 \end{equation}
where the second equivalence obviously ignores potential phases or anomalies and we are not being explicit in coupling-dependent Jacobians (see below). 

%
  The values of the discretised (scalar) fields are assigned to each lattice site,  with periodic boundary conditions for all the fields except for antiperiodic temporal boundary conditions in the case of fermions. The discrete approximation of continuum derivatives are finite difference operators defined on the lattice. A Wilson-like lattice operator must be introduced~\cite{toappear}, such that   fermion doublers are suppressed 
 and the one-loop constant $-3\ln 2/\pi$  in (\ref{cuspperturbative}) is recovered in lattice perturbation theory.  
While such Wilson term explicitly breaks the $SO(6)$ symmetry of the model, for  the $SO(6)$ singlet  quantities that we study -- $f(g)$ and $x,x^*$ correlators -- this might affect  the way one approaches the continuum limit, an issue which we address in the next Section. 

The Monte Carlo evolution of the action \eqref{Scuspquadratic}   is generated by the standard Rational Hybrid Monte Carlo (RHMC) algorithm,  
 as in \cite{Roiban}. A detailed analysis of the precision of the fourth-root approximation and the magnitude of the fermion matrix eigenvalues \eqref{OF}, error analysis and auto-correlation times of the Monte Carlo histories is in \cite{toappear}.

\section{Simulation, continuum limit and the phase}
\label{sec:simulation}
As discussed above, in the continuum model there are two parameters,  the  dimensionless coupling $g=\frac{\sqrt{\lambda}}{4\pi}$  and the mass scale $m$. 
In taking the continuum limit, the dimensionless physical quantities that it is natural to keep constant are the physical masses of the field excitations  rescaled by $L$, the spatial lattice extent. This is our line of constant physics.  For the example in \eqref{mx}, this means 
\begin{equation}\label{constantphysics}
L^2 \,m^2_{x}=\text{const}\,,\qquad~~\text{which leads to}\qquad~~L^2 \,m^2\equiv (N M)^2 =\text{const}\,,
\end{equation}
where we defined the dimensionless $M=m a$ with the lattice spacing $a$.  
The second equation in \eqref{constantphysics} relies first on the assumption that $g$ is \emph{not} renormalized, which is suggested  lattice perturbation theory~\cite{toappear}. Second, one should investigate whether   relation \eqref{mx},  and the analogue ones for the other fields  of the model, are still true in the discretized model - \emph{i.e.} the physical masses undergo only a \emph{finite} renormalization. In this case, at each fixed $g$ fixing $ L^2~m^2$ constant would be enough to keep the rescaled physical masses constant, namely   no tuning of the ``bare'' parameter $m$ would be necessary.  
\begin{figure}[t]
    \centering
               \includegraphics[scale=0.65]{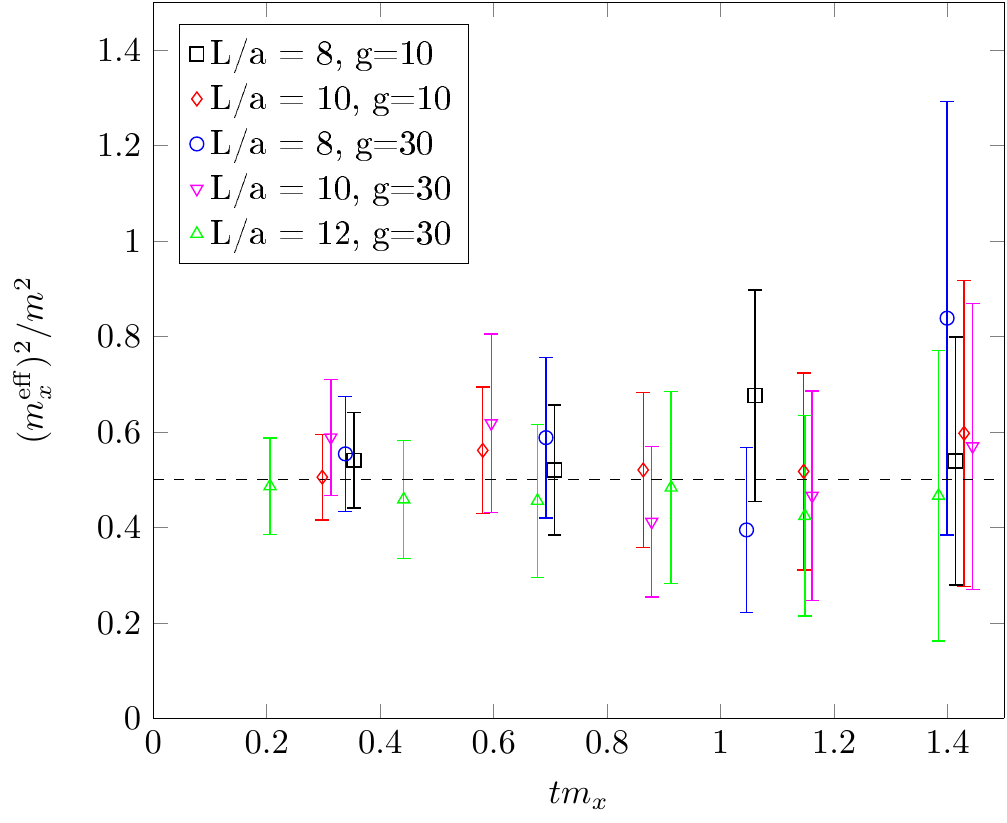}
              \hspace{1.5cm}
               \includegraphics[scale=0.65]{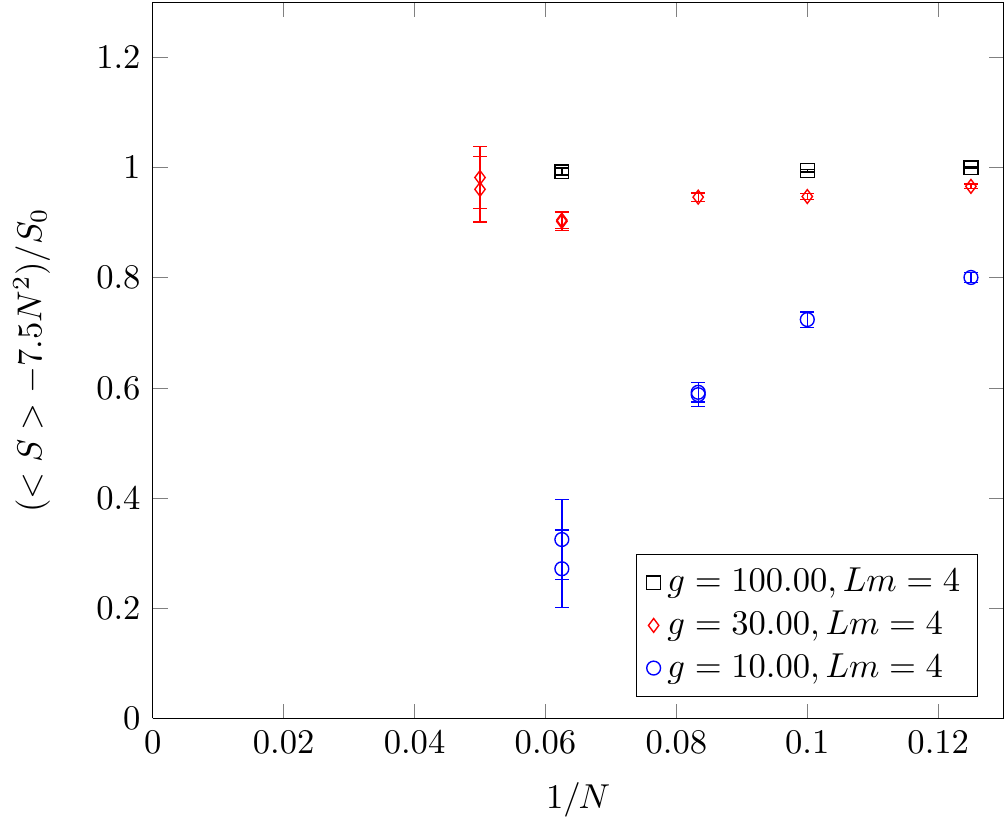}
        \caption{{\bf Left panel:} Effective mass plot $m^{\rm eff}_x=\frac{1}{a}\ln\frac{C_x(t)}{C_x(t+a)}$, as calculated from the  correlator $C_x(t)=\sum_{s_1,s_2} \langle x(t,s_1) x^*(0,s_2)\rangle$ of bosonic fields $x,x^*$ in  presence  of Wilson terms. \\
         {\bf Right panel:} Plot of the ratio $\frac{\langle S_{LAT}\rangle-c\, N^2/2}{S_0} $, where $S_0=M^2\,N^2\,g/2$ and  $c/2=7.5(1)$.} 
     \label{fig:corr&action}
\end{figure}
In this contribution, we start by considering the example of bosonic $x,x^*$ correlators, where indeed we find no ($1/a$) divergence for the ratio $m_x^2/m^2$ -- see the left panel in Figure \ref{fig:corr&action}. In the  large $g$  region that we investigate  
the ratio considered approaches the expected continuum value $1/2$.   
Having this as hint, and because with the proposed discretization we recover in perturbation theory the one-loop cusp anomaly~\cite{toappear}, we assume  that in the discretized model no further scale but the lattice spacing $a$  is present.  Any  observable $F_{\rm LAT}$ is therefore a function $F_{\rm LAT}=F_{\rm LAT}(g,N,M)$ of the  input (dimensionless) parameters $g=\frac{\sqrt{\lambda}}{4\pi}$, $N=\frac{L}{a}$ and  $M=a \,m$.
In Table \ref{t:runs}  we list the parameters of the simulations presented in this paper. 
 At fixed coupling $g$ and fixed $m\, L\equiv M\,N$ (large enough so to keep finite volume effects $\sim e^{-m\,L}$  small),    $F_{\rm LAT}$ is evaluated for different values of $N$. The continuum limit  -- which we do not attempt here -- is then obtained extrapolating to infinite~$N$.
%
%

In measuring  the action \eqref{vevaction} on the lattice, we are supposed to recover the following general behavior 
\begin{equation} \label{fit}
\frac{\langle S_{\rm LAT}\rangle}{N^2} = \frac{c}{2}+\frac{1}{8}\,M^2\,g \,f'(g)\,, 
\end{equation}
 where we have reinserted the parameter $m$, used  that $V_2=a^2\,N^2$ and added a constant  contribution  in $g$ which takes into account possible coupling-dependent Jacobians relating the (derivative of the) partition function on the lattice to the one in the continuum. 
Measurements for the ratio
$ 
\frac{\langle S_{LAT}\rangle-c\, N^2/2}{M^2\,N^2\,g/2} = \frac{f'(g)}{4} ~
$
 are, at $g=100$,  in good agreement with $\frac{c}{2}=7.5(1)$ -- consistently with the expectation~\cite{toappear}
-- and with the leading order prediction in  \eqref{cuspperturbative} for which $ f'(g) =4$.   For lower values of $g$ -- red and blue curves in Figure \ref{fig:corr&action}, right panel -- we observe a deviation that obstructs the continuum limit and is compatible with the presence of a divergence.  This could be explained with the presence of infinite mass renormalization in those fields correlators which we have not considered so far, and whose investigation is left for the future.
 
\begin{figure}[h]
   \centering
 \includegraphics[scale=0.35]{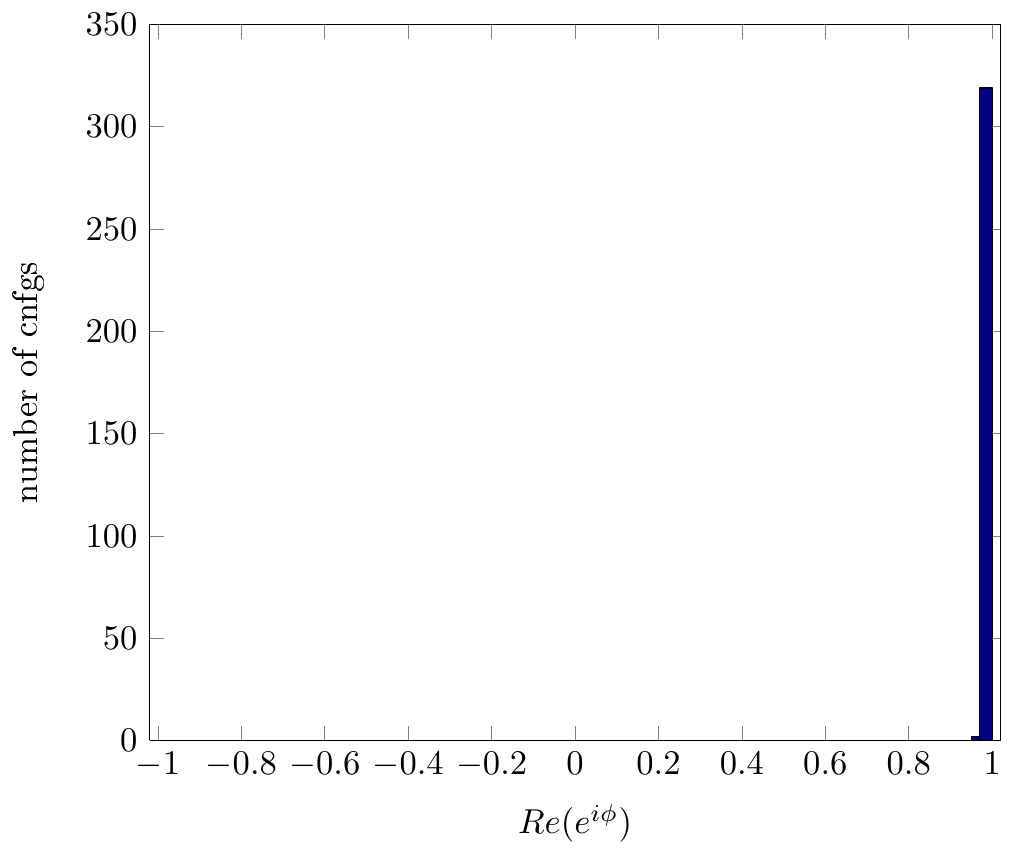}
  \includegraphics[scale=0.35]{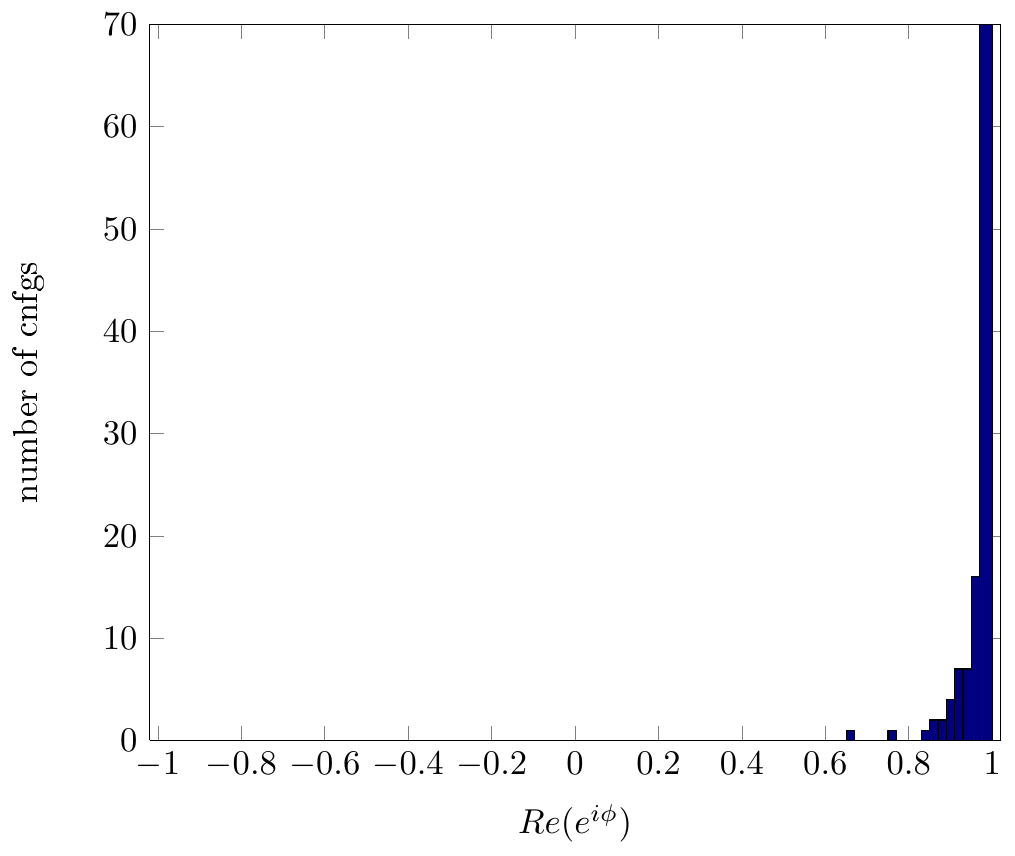}
   \includegraphics[scale=0.35]{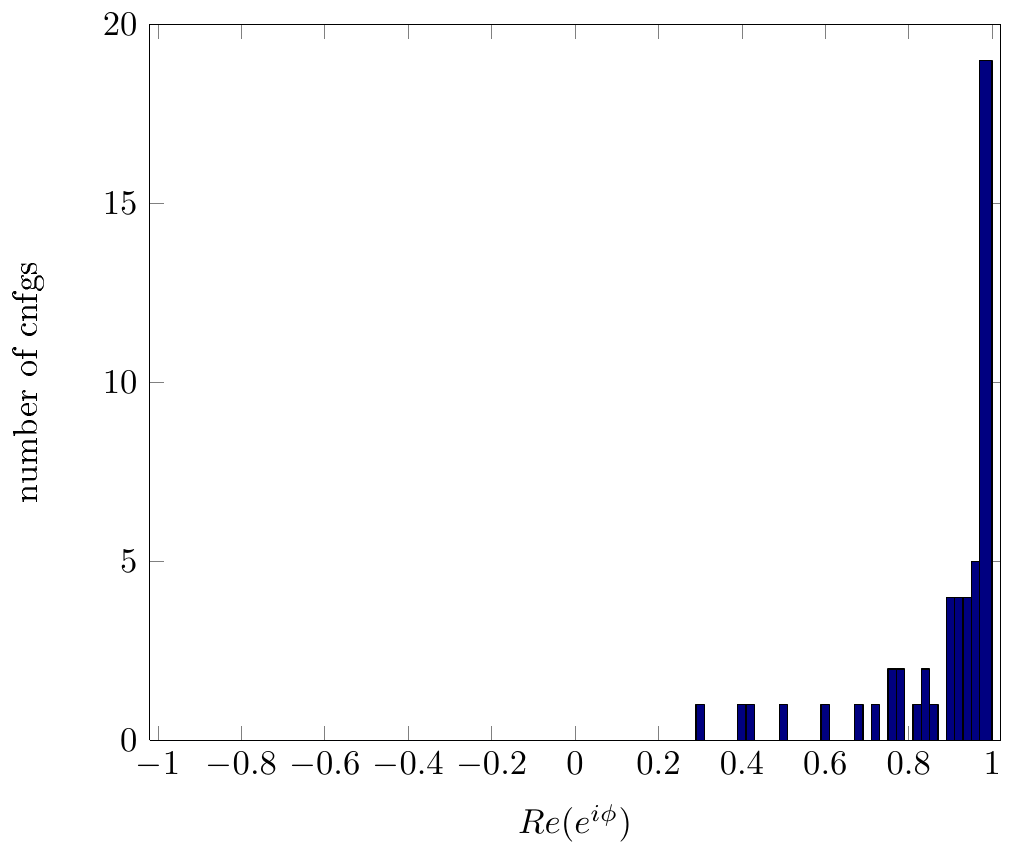}
    \includegraphics[scale=0.37]{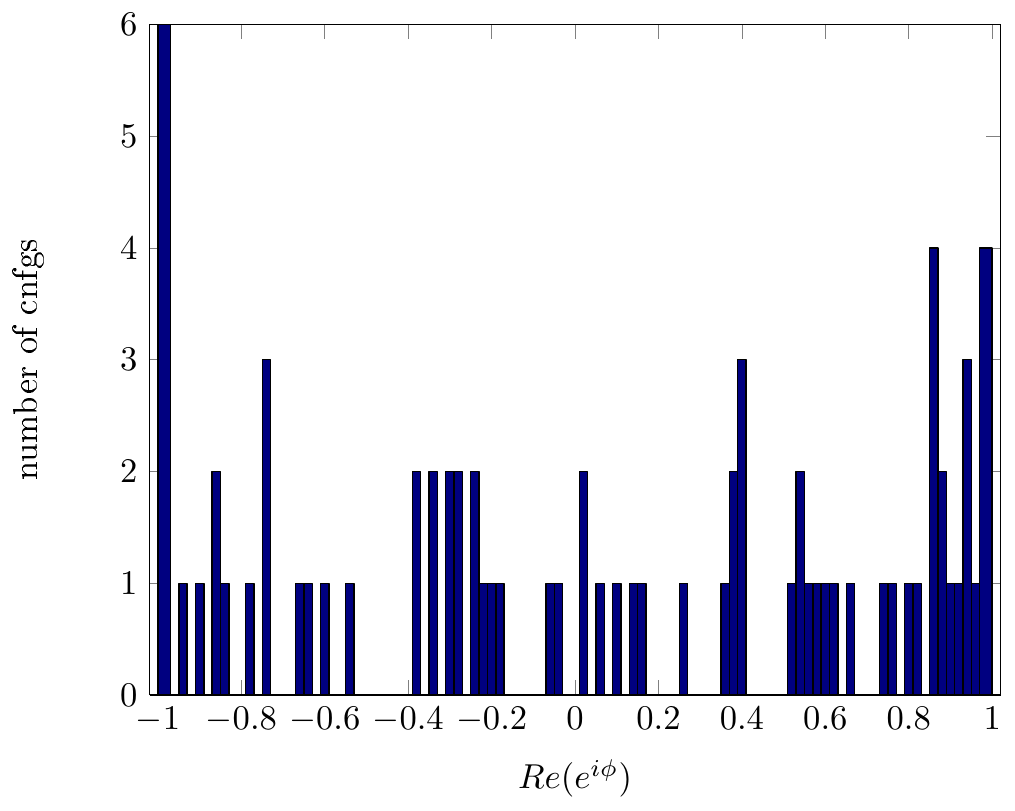}
 \caption{Histograms for the  frequency of the real part (the imaginary part vanishes by symmetry) of the   reweighting phase factor $e^{i\theta}$ of the Pfaffian ${\rm Pf}\,O_F=|(\det O_F)^{\frac{1}{2}}|\,e^{i\theta}$, based on the ensembles generated  at  $g=30,10,5,1$ (from left to right). At $g=1$ the observed $\langle e^{i\theta}\rangle$ is consistent with zero, thus  preventing the use of standard reweighting. }
\label{fig:phase}
\end{figure}

In proximity to $g\sim 1$,  
severe  fluctuations appear in the averaged phase of the Pfaffian    -- see Figure \ref{fig:phase} -- signaling the sign problem mentioned in the Discussion. 
%

\begin{table}[t]
 \centering
\begin{tabular}{cccccc}
\toprule
$g$ & $T/a\times L/a$ & $am$ & $Lm$ & statistics [MDU] & $\tau_{\rm int}$ \\
\midrule
10    & $16\times  8$ & 0.5  & 4 & 900 & 2 \\
     & $20\times 10$ & 0.4  & 4 & 900 & 3 \\
     & $24\times 12$ & 0.33333  & 4 & 821, 900 & 5 \\
     & $32\times 16$ & 0.25  & 4 & 316, 357 & 10 \\
\midrule
30    & $16\times  8$ & 0.5  & 4 & 800 & 1 \\
     & $20\times 10$ & 0.4  & 4 & 800 & 2 \\
     & $24\times 12$ & 0.33333  & 4 & 900 & 4 \\
     & $32\times 16$ & 0.25  & 4 & 625, 800 & 8 \\
     & $40\times 20$ & 0.2  & 4 & 300, 300 & 60 \\
\midrule
100   & $16\times  8$ & 0.5  & 4 & 900 & 1 \\
     & $20\times 10$ & 0.4  & 4 & 750 & 2 \\
     & $32\times 16$ & 0.25  & 4 & 411, 415 & 5 \\
\bottomrule
\end{tabular}
 \caption{
 Parameters of the simulations. The temporal extent $T$ is always twice the spatial extent $L$, which helps studying the correlators.
 The size of the statistics after thermalization is given in terms of Molecular Dynanic Units (MDU) which equal an HMC trajectory of length one.
 In the case of multiple replica the statistics for each replica is given. The typical auto-correlation time 
 of the correlators is given in the last column.
}
 \label{t:runs}
\end{table}

%
\vspace*{-0.9em}
\acknowledgments
\vspace*{-0.8em}

\noindent We are particularly grateful to M. Bruno for initial collaboration, and to R. Roiban, D. Schaich, R. Sommer and A. Wipf  for very useful elucidations. We also thank F. Di Renzo, H. Dorn, G. Eruzzi, M. Hanada and the theory group at Yukawa Institute, B. Hoare, B. Lucini, J. Plefka, A. Schwimmer, S.  Theisen, P. T\"opfer, A. Tseytlin and U. Wolff for useful discussions.
 
\appendix

\end{document}